\documentclass[a4paper]{jpconf}
\usepackage{graphicx}
\usepackage{amsmath}
\usepackage{slashed}
\usepackage{cite}

\newcommand{\beq}{\begin{eqnarray}}
\newcommand{\eeq}{\end{eqnarray}}

\newcommand \spinhalf {spin-$\frac{1}{2}$~}
\newcommand \mpcac {m_{\mathrm{PCAC}}}

\newcommand{\TR}{\mathrm{Tr}}
\newcommand{\RE}{{\cal R}\mathrm{e}}
\newcommand{\dslash}{\slashed{\partial}}

\begin{document}
\title{Numerical results for gauge theories near the conformal window}

\author{Biagio Lucini}

\address{Physics Department, College of Science, Swansea University,
  Singleton Park, Swansea SA2 8PP, UK}

\ead{b.lucini@swansea.ac.uk}

\begin{abstract}
A novel strong interaction beyond the standard model could provide a
dynamical explanation of electroweak symmetry breaking. Experimental
results strongly constrain properties of models that realise this
mechanism. Whether these constraints are obeyed by any strongly
interacting quantum field theory is a non-perturbative problem that
needs to be addressed by first-principle calculations. Monte Carlo
simulations of lattice regularised gauge theories is a powerful tool
that enables us to address this question. Recently various lattice
investigations have appeared that have studied candidate models of
strongly interacting dynamics beyond the standard model. After a brief
review of the main methods and of some recent results, we focus on the
analysis of SU(2) gauge theory with one adjoint Dirac fermion flavour,
which is shown to have a near-conformal behaviour with an anomalous
dimension of order one. The implications of our findings are also
discussed. 
\end{abstract}

\section{Introduction and motivations}
The recent experimental discovery of the Higgs
boson~\cite{Aad:2012tfa,Chatrchyan:2012ufa} is one of the most
remarkable confirmations of the validity of the standard model (SM) of
particle physics interactions, which embodies electroweak symmetry
breaking through the presence of a 
condensate for the Higgs field. However, despite the spectacular success of this
theory (leading for instance to the most accurate predictions for the
fine constant structure $\alpha$~\cite{Gabrielse}), from the
theoretical point of view there are clear indications that the
SM (or more precisely the Higgs sector of the
SM) is not a fundamental theory. One of the problems with
the SM Higgs mechanism is related to fine tuning: due to loop
contributions, the natural order of magnitude for the mass of the
Higgs boson is the mass of the cut-off of the theory (which,
ultimately, is the Planck scale $m_P \simeq 10^{19}$ GeV), while
experimentally it is found that the Higgs mass $m_H$ is around 125
GeV.  If the SM were to be a fundamental theory, such a low
value for $m_H$ could only result from very precise cancellations that
span seventeen orders of magnitude. 

Although the scenario of the SM as a fundamental theory could
be related to asymptotic safety in gravity~\cite{Shaposhnikov:2009pv},
a huge cancellation is generally regarded as the manifestation of
the fact that the theory is either incomplete or not
fundamental. For instance,
the mass of the scalar is protected from acquiring values of the order
of the cut-off scale if the theory is supersymmetric, with
contributions of bosonic and fermionic particles cancelling each
others in the self energy of the Higgs. Another possibility is that
the Higgs boson be a composite particle resulting from a new strong
interaction. This scenario is referred to as {\em strongly interacting
  dynamics beyond the standard model} or more commonly as {\em
  technicolor}. The basic idea of this framework is that the
electroweak symmetry is broken spontaneously by a chiral condensate
of a new strong dynamics involving particles that are not in the
SM~\cite{Weinberg:1975gm,Susskind:1978ms,Eichten:1979ah}. This 
mechanism of electroweak symmetry breaking is lifted from Quantum
Chromodynamics (QCD), in which the quark
chiral condensate does break electroweak symmetry, although the
phenomenology of this breaking is three orders of magnitude weaker than
required in the SM. Although a simple copy of QCD with a
larger $\Lambda$ parameter is incompatible
with electroweak precision measurements~\cite{Peskin:1991sw}, it is
possible to reconcile dynamical electroweak symmetry breaking with
experiments if one assumes that the running of the gauge coupling of
the novel interaction is slower than in QCD~\cite{Holdom:1984sk,Appelquist:1986an,Yamawaki:1985zg}. This happens if the theory
is close to the conformal window. In this case, approximate scale
invariance in the infrared (known as near-conformality) determines the
appearance of a light scalar particle in the
spectrum~\cite{Yamawaki:1985zg} that can play the role of the SM
Higgs boson. The basic ideas of technicolor are reviewed for
instance in~\cite{Hill:2002ap}. 

Determining the dynamical features of technicolor-like theories is a
non-perturbative problem. As such, it requires approaches that are
controlled at strong coupling. In this respect, much progress has
been recently achieved using string-gauge duality techniques, 
starting from first-principle top-down
approaches~\cite{Nunez:2008wi,Elander:2009pk,Nunez:2009da,Elander:2011mh,Conde:2011aa,Conde:2011ab,Elander:2012yh,Elander:2014ola},
as well as in the probe brane approximation~\cite{Carone:2007md,Hirayama:2007hz,Mintakevich:2009wz,Anguelova:2010qh,Anguelova:2011bc,Anguelova:2012ka,Kutasov:2012uq,Hoyos:2013gma,Faedo:2013ota,Faedo:2014naa} and utilising
bottom-up
techniques~\cite{Hong:2006si,Hirn:2006nt,Carone:2006wj,Fabbrichesi:2008ga,Haba:2008nz,Hirn:2008tc,Round:2010kj,Alho:2013dka,Evans:2014nfa,Erdmenger:2014fxa}. A
valid alternative to analytic treatments is a numerical 
approach based on the framework of Lattice Gauge Theories. Originally
devised for studying low energy phenomena in QCD, the lattice can be
proficiently used also for studying strongly interacting dynamics
beyond the SM. Since the physics in this case is
inherently different from that of QCD, a novel set of numerical tools
has been developed to investigate gauge theories near the conformal
window and to answer the different type of questions that arise in this
context. 

In only few years after the original lattice studies of dynamical
electroweak symmetry breaking due to a novel
interaction~\cite{Catterall:2007yx,Appelquist:2007hu,Shamir:2008pb,DelDebbio:2008zf,Deuzeman:2008sc},
the field of lattice investigations of near-conformal gauge theories
has made an enormous progress in answering crucial questions
related to the theoretical viability of strongly interacting dynamics
beyond the SM. The purpose of this work is to provide an
introduction to the field and some of the related questions that can
be meaningfully answered with lattice simulations contextualised with
a few examples. For a comprehensive review that also make better
justice to the wide literature in the field, we refer for instance
to~\cite{Giedt:2012it,Kuti:2014epa,DelDebbio:2014ata}.

The rest of this article is organised as follows. In
Sect.~\ref{sect:2}, we introduce the framework of dynamical electroweak
symmetry breaking and justify the need for lattice
calculations. An overview of lattice techniques is provided in
Sect.~\ref{sect:3}. Sect.~\ref{sect:4}~and~Sect.~\ref{sect:5} report
on numerical results for two gauge theories (respectively SU(2) with
two adjoint Dirac fermions and SU(2) with one adjoint Dirac fermion)
that can be used to advance our understanding of dynamical
electroweak symmetry breaking due to a new strong force. Finally,
Sect.~\ref{sect:6} presents concluding remarks and gives an overview
of future directions.

\section{Novel strong dynamics and electroweak symmetry breaking}
\label{sect:2}
The SM is a $SU(2)_L \otimes U(1)_Y$ gauge theory coupling
doublets of left-handed fermions to four gauge bosons. 
In addition to fermionic matter and gauge bosons, the SM
contains a  doublet of scalars with a quartic self-interaction
potential having minima at a non-perturbative vacuum.
The scalar field gets a non-trivial vacuum expectation value ({\em
  vev}), reducing the gauge group to $U(1)_{EM}$ and providing mass
to three gauge bosons. Besides, fermions get mass from the
Higgs {\em vev} via a Yukawa interaction. This picture has been
confirmed to an extraordinary degree of accuracy by electroweak
precision measurements performed at LEP and at Tevatron and by the
recent discovery of the Higgs boson at the LHC.

However, from a more fundamental point of view, the picture is still
unsatisfactory. One of the key questions is the following. Due to
quantum fluctuations, the Higgs mass is expected to get corrections of
the order of the natural cut-off (Planck scale); what does keep it at
around 125 GeV? The answer to this question is not unique. Elegant and
appealing scenarios can be conjectured if one considers the SM as an
effective theory that is only valid at scales below the TeV. Following
this route, the Higgs field is complemented with or replaced by a whole new
sector. This sector should provide mass to the weak interaction gauge
bosons and to the SM fermions while at the same time being
compatible with the very stringent phenomenological constraints. The recent
breakthrough at the LHC imposes the presence of a light scalar, the
state that in the SM is identified with the Higgs boson. 

Among possible scenarios, one of the proposed possibilities is a new
interaction that couples new fermions to new gauge bosons. In QCD,
chiral symmetry is spontaneously broken by the 
formation of a chiral condensate, which is not invariant under
$SU(2)_L \otimes U(1)_Y$. Hence, the strong force by itself provides
electroweak symmetry breaking. However, the SM strong
force is not able to account for the phenomenologically observed
electroweak symmetry breaking of the SM, its typical
infrared (IR) scale being $\Lambda_{QCD} \simeq 200$ MeV, since in the SM,
electroweak symmetry breaking is characterised by an intrinsic scale
of about $250$ GeV. Hence, going down the strong interaction route to
explain electroweak symmetry breaking, we need a new force.
Mass to the weak gauge bosons is provided through this strong dynamics
as follows. As a consequence of spontaneous symmetry breaking, a set of Goldstone
bosons emerge.  Among those bosons, three provide the
longitudinal component to the $Z$ and $W^{\pm}$ bosons and the others
acquire mass of the order of the strong scale of the theory (around one
TeV). In order to provide mass to the SM fermions, a further
non-Abelian interaction is required that connects these fermions with
the fermions of the new force.

Despite the appeal of this framework and the comfort that we have a deep
understanding of chiral symmetry breaking, assuming a QCD-like new
strong force generates tension between the need to provide mass to
fermions and suppression of flavour changing
neutral currents. The tension can be mitigated and hopefully removed if the
theory does not have a standard QCD-like dynamics, but is instead
characterised by a dynamics with two scales. At the larger scale,
$\Lambda_{UV}$, the physics would be perturbative and dominated by the
Gaussian fixed point. At scales below $\Lambda_{IR}$, the smaller
scale, the dynamics will be chiral symmetry breaking. In an
intermediate regime, the coupling might in principle evolve slowly
(i.e. it could {\em walk} instead of
running)~\cite{Holdom:1984sk,Appelquist:1986an,Yamawaki:1985zg}. Yet
another requirement is the anomalous dimension $\gamma_{\star}$ of the
chiral condensate being very close to one~\cite{Chivukula:2010tn}. 

Walking can be achieved by adjusting the parameters of the theory (number
of flavours and of colours) so that the model is still confining, but
in the vicinity of the onset of the conformal window. It is worth at
this point to remind the possible IR behaviours (phases) of a non-Abelian
gauge theory. In the perturbative region, the evolution of the
gauge coupling $g$ with the renormalisation scale $\mu$ is given by
\begin{eqnarray}
\beta(g^2) = \mu \frac{d g}{d \mu}  = - b_0 g^3 - b_ 1 g^5 + \dots \ ,
\end{eqnarray}
and the two universal (i.e. renormalisation scheme independent)
coefficients $b_0$ and $b_1$ can be computed in perturbation
theory. For a $SU(N)$ gauge theory with  $N_f$ flavours of massless fermions in the
representation $R$ of the gauge group, one obtains
\begin{eqnarray}
 b_0 = \frac{1}{(4 \pi^2)} \left( \frac{11}{3} N - \frac{4}{3} T_R N_f
\right) \ , \qquad b_1 = \frac{1}{(4 \pi)^4} \left[\frac{34}{3} N^2 -
  \frac{20}{3}N T_R N_f- 4 \frac{N^2 - 1}{d_R} N_f  \right] \ ,
\end{eqnarray}
where $T_R$ and $d_R$ are
respectively the normalisation of the trace and the dimension of the
representation $R$. Physically interesting theories are those that
display asymptotic freedom. This property implies $b_0 > 0$. In QCD,
also $b_1$ is bigger than zero, and this gives rise to confinement and
chiral symmetry breaking. However, for particular choices of $N_f$ and
of $R$, it is possible to obtain $b_1 < 0$ while retaining asymptotic
freedom. This gives rise to a zero in the beta function $\beta(g^2)$, which
determines the existence of a fixed point in the IR. For the
scenario to be consistent, one needs the fixed point coupling $g_{\ast}$ to be
small.  A similar IR fixed point is known as the Caswell-Banks-Zaks fixed
point~\cite{Caswell:1974gg,Banks:1981nn}. It is believed (and
confirmed in supersymmetry, where analytic calculations are possible) 
that IR conformality survives also for values of $N_f$ at which
the coupling is stronger, up to a lower value $N_f^{c,l}$ at which the
dynamics becomes confining and chiral symmetry breaking like in
QCD. At fixed $N$, the values of $N_f$ such that $N_f^{c,l} \le N_f
\le N_f^{c,u}$, where the upper value corresponds to the loss of
asymptotic freedom, identifies theories with an IR fixed
point. This region is called the {\em conformal window}. 

\begin{figure}
\begin{center}
\begin{tabular}{ccc}
\includegraphics[width=.33\textwidth]{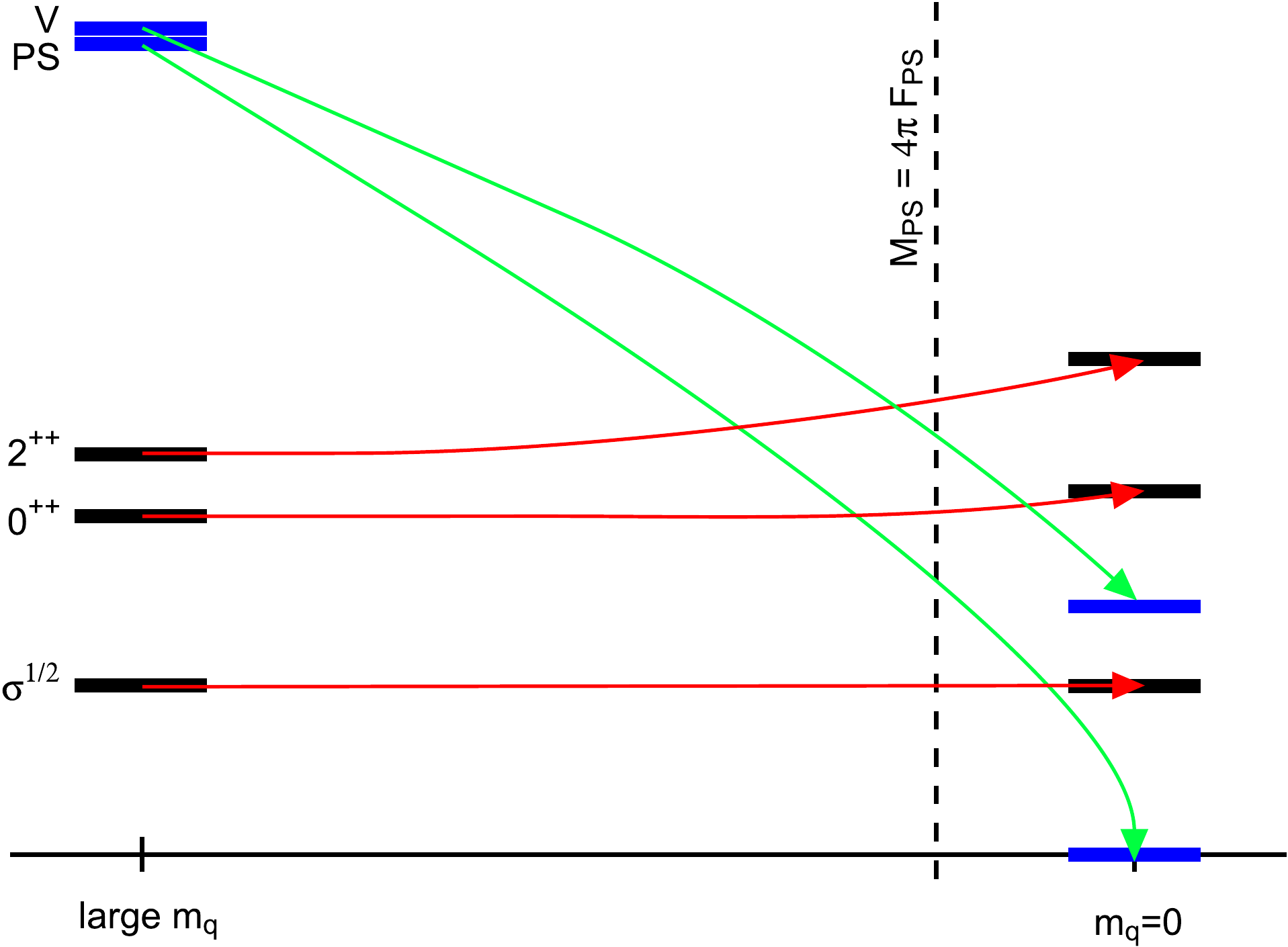} & 
\includegraphics[width=.33\textwidth]{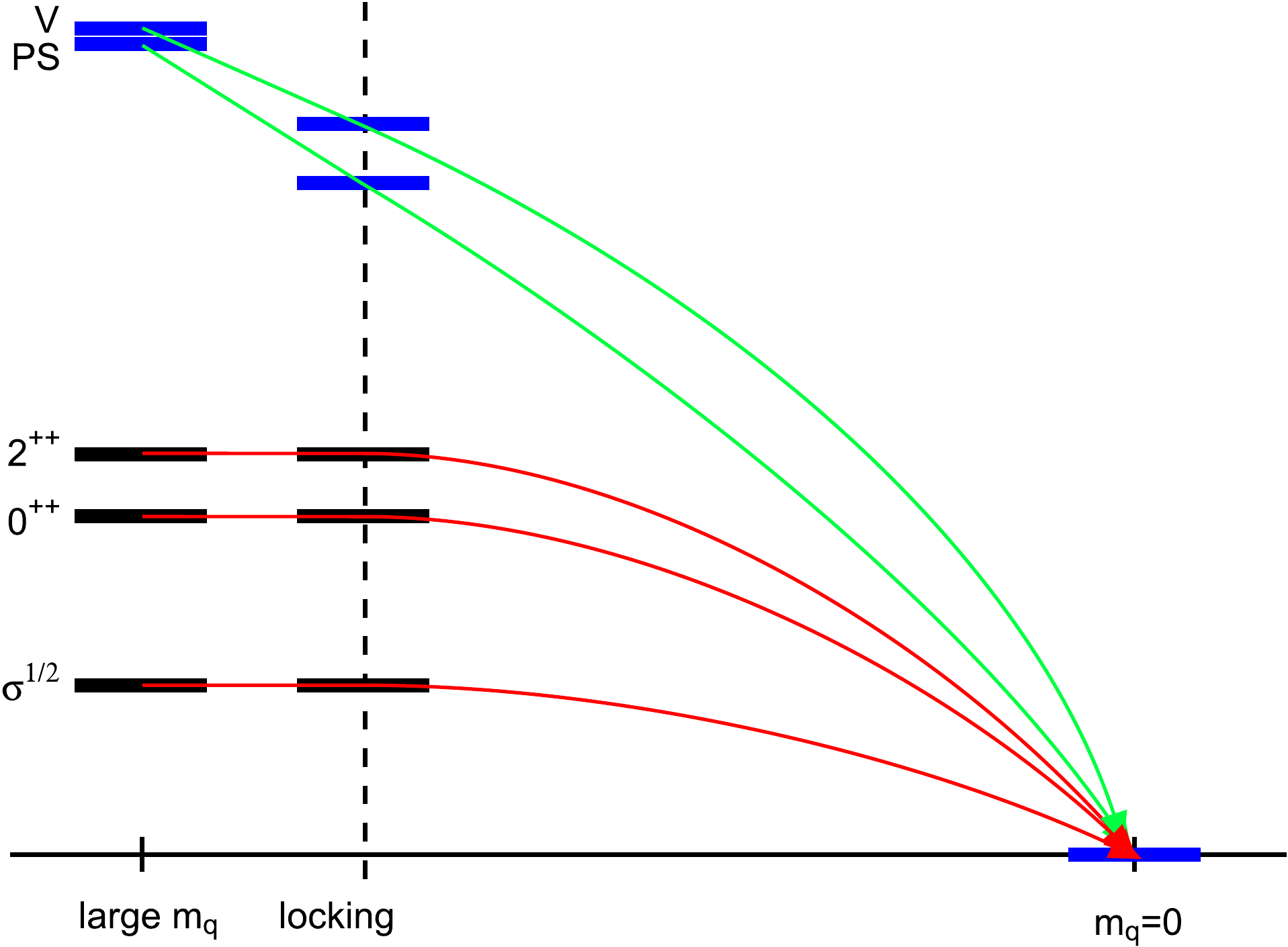} &
\includegraphics[width=.33\textwidth]{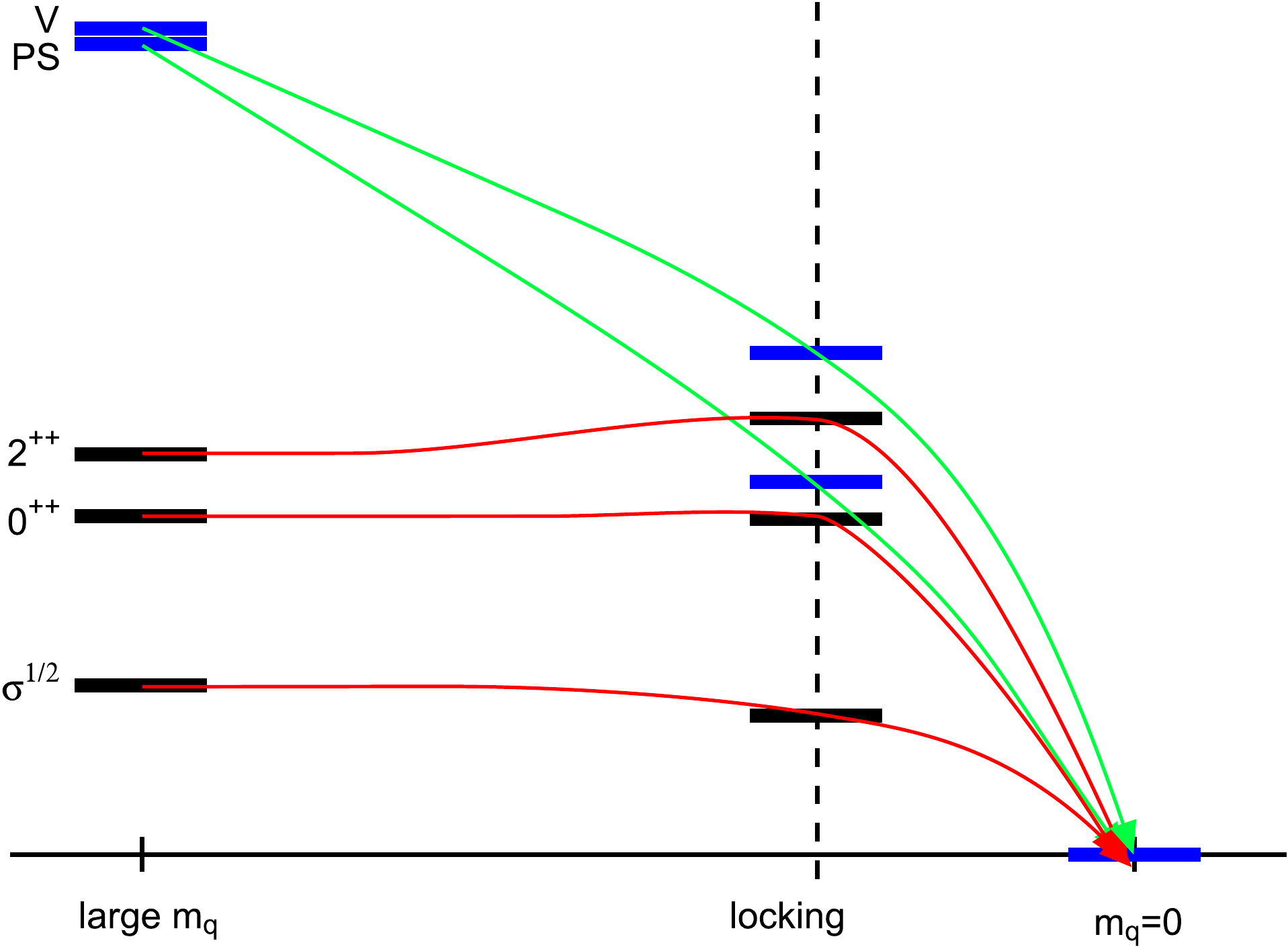}\\
(a) & (b) & (c)
\end{tabular}
\end{center}
\caption{Behaviour of a SU($N$) gauge theory with fermionic matter
as the fermion mass is lowered to zero from the high-mass regime  when
(a) the system has a QCD-like IR dynamics; (b) the system is IR conformal, with
locking occurring in the high-mass regime; (c) the system is
IR conformal, with locking occurring at or below the scale of
chiral symmetry breaking.}
\label{fig:lock}
\end{figure}

Since walking can happen below the onset of the conformal window, the
problem is characterised by a strong coupling and requires an
investigation method that is controlled in the non-perturbative
regime. Below, we shall use lattice regularisation followed by
Monte Carlo simulations. In order for numerical methods to work, a
non-zero fermion mass is needed. Results for the massless limit can then
be obtained by extrapolating data for non-zero fermion mass to the
chiral point. In order to understand lattice results, in the remainder
of this section we  review what happens to IR conformal gauge
theories when they are deformed with a small mass term. We start by
reviewing the behaviour of the QCD spectrum as the fermion mass $m$ is
varied. At $m = \infty$, mesons decouple and we have a Yang-Mills
spectrum. If the mass $m$ is lowered, we are in a heavy quark regime,
in which mesons are more massive than glueballs and scalar and
pseudoscalar mesons are nearly degenerate. If we lower the mass, when we hit the
chiral scale, the pseudoscalars will start to emerge as the
pseudo-Goldstone bosons of chiral symmetry breaking and get lighter,
while the rest of the spectrum will settle at the QCD scale. Finally,
in the chiral limit the pseudoscalars become massless. This is
sketched in Fig.~\ref{fig:lock}a. 

In the massless case, for a theory with an IR fixed point, if
we assume a regular behaviour for the renormalisation group (RG) functions, we find 
\begin{eqnarray}
g \to g_\ast: \ 
\left\{
\begin{array}{l}
\beta(g) \simeq \beta_\ast ( g - g_\ast ) \\ 
\gamma(g) \simeq \gamma_\ast  
\end{array}
\right. \ ,
\end{eqnarray}
where $\gamma_\ast$ is the fixed point value of the chiral condensate
anomalous dimension $\gamma(g)$. 
In order to understand what happens when we deform the theory with a
mass term, we remark that the mass is a relevant direction in the RG
sense. As a consequence, any non-zero mass term determines the loss of
IR conformality and the appearance of a confining and chiral
symmetry breaking spectrum. The running of the mass with the
renormalisation scale $\mu$ from a reference scale $\mu_0$ is given by
\begin{equation} \nonumber
m(\mu) = m(\mu_0) \exp \left\{ - \int_{g(\mu_0)}^{g(\mu)}
  \frac{\gamma(z)}{\beta(z)} dz \right\} \equiv Z(\mu,\mu_0,\Lambda)
m(\mu_0) \ .
\end{equation}
We can define a renormalised mass $M$ from the condition $m(M) = M$. A
large value of $M$ destroys conformality and the theory looks like
Yang-Mills with heavy matter . In this case, the mesons will have
mass $m_{mes} \simeq 2 M$ and the glueballs $m_{glue} \simeq
B_{glue} \Lambda$, with $\Lambda$ the Yang-Mills scale and $B_{glue}$ some numeric
coefficient not necessarily small (in QCD, for instance, the mass of
the lightest glueball is about eight times $\Lambda_{QCD}$). We are
interested in the opposite regime, in which $M \ll \Lambda$. An
analytic calculation can be performed near the Caswell-Banks-Zaks
fixed point~\cite{Miransky:1998dh}. The central result is the
emergence of an IR scale $\Lambda_{IR}$ related to $M$ but exponentially
suppressed with respect to it:
\beq
\Lambda_{IR} = M \ e^{ - \frac{1}{2 b_0^{YM} g^2_\ast } } \ll M \ll
\Lambda \ .
\eeq
 $\Lambda_{IR}$ controls an effective large-distance Yang-Mills spectrum, with
the fermions being in a heavy quark regime. At energies much lower
than $M$, the spectrum is that of a pure Yang-Mills theory with scale
$\Lambda_{IR}$. Mesons are bound states of the quark-antiquark pairs interacting
via the YM static potential, the bound energy is small with respect to
the mass of the fermions, and the correction to the potential due  to
quark-antiquark pair creation are negligible. As a result, mesons are
effectively quenched and have a mass that is much larger than the mass
of the lowest-lying glueballs. As the mass $M$ is reduced, the IR
physics is always the same,  provided that all the masses are
rescaled with $M$.  

One can show that spectrum of a particle
$X$ varies with the constituent mass $m$ and with the anomalous dimension
of the condensate at fixed point as follows: 
\begin{equation}
m_X = A_X \mu^{\frac{\gamma_\ast}{1+\gamma_\ast}} m(\mu)^{\frac{1}{1+\gamma_\ast}} \ .
\end{equation}
On a lattice of spacing $a$, choosing $\mu=a^{-1}$ gives
\begin{eqnarray}
\label{eq:scaling}
a m_X = A_X (a m)^{\frac{1}{1+\gamma_\ast}} \ ,
\end{eqnarray}
with $A_X$ independent of $m$ (at the leading order, in the chiral
region). As a consequence, ratios of masses are constant as a function of
$m$. Note that this is in stark contrast with QCD, for which the ratio
of the mass of any confining state over the pseudoscalar mass diverges in the
chiral limit. 

In summary, for a Caswell-Banks-Zaks fixed point, one expects that for
large constituent masses the scenario is still that of Yang-Mills plus
heavy quarks. As the mass deformation regime is reached, the spectrum
will scale uniformly towards the chiral limit, keeping constant mass
ratios. This locking of the spectrum arises in the heavy quark
regime (Fig.~\ref{fig:lock}b), which will be a feature of the
mass-deformed theory regardless of how small the mass deformation is. 

If we leave the perturbative fixed point, analytic predictions are not
possible. However, we still expect that the general feature of the
perturbative description (i.e. constant spectral ratios) characterises the mass deformed
regime, but the locking will happen when the spectrum of the large
distance theory has left the heavy quark regime. In particular,
we might expect that for this IR theory the pseudoscalar mesons
emerge as pseudo-Goldstone bosons related to chiral symmetry
breaking, as in QCD, or, more interestingly, they can have a mass that
is comparable to that of the lightest glueball  (Fig.~\ref{fig:lock}c).

In the context of our discussion of confining and chiral symmetry
breaking gauge theories versus IR conformal gauge theories,
walking can be seen as a cross-over phenomenon: a theory is walking
if for an intermediate regime of energies it presents features that one
would attribute to an IR conformal gauge theory, but as the
energy is reduced, eventually becomes confining and chiral symmetry breaking.
Because of the behaviour at intermediate scales, walking is also
referred to as {\em near-conformality}.

Since the analysis we have performed so far is only semi-quantitative, our
discussion leaves unanswered important fundamental questions,
starting from whether this scenario is realised at all in a concrete
gauge theory. Those issues will be addressed via Monte Carlo
calculations on the lattice.

\section{Lattice setup}
\label{sect:3}
Before discussing our results, we briefly review how gauge theories
are formulated on a lattice. The advantage of the lattice formulation
over the continuum one is that the former is amenable to Monte Carlo
numerical simulations, which  can be used to study non-perturbative
features of the theory. 

Consider a gauge theory with gauge group ${\cal G} = SU(N)$ and $N_f$
fermionic fields  in the representation $R$ of ${\cal G}$. A
formulation of the theory that preserves gauge invariance on a
Euclidean four-dimensional grid with an isotropic spacing $a$ can be obtained as
follows. If $A_{\mu}$(x) is the vector potential, we define the link
variable $U_{\mu}(i)$ on the grid point $i = x/a \equiv (i_0, \dots, i_3)$ as
\beq
\label{eq:link}
U_{\mu}(i)= \mathrm{P exp} \left( i g \int_{a i} ^{a(i + \hat{\mu})} A_{\mu}(x) \,
  \mathrm{d} x^{\mu}\right) \ ,
\eeq
with $\hat{\mu}$ the versor in direction $\mu$ and $g$ the gauge
coupling. $U_{\mu}(i)$ is naturally associated with the link $(i;
\hat{\mu})$ that joins the point $i$ with the point $i +
\hat{\mu}$. The variable corresponding to the negative direction $-\mu$ is given by
\beq
U_{-\mu}(i)= \left(U_{\mu}(i) \right)^{\dag} \equiv U^{\dag}_{\mu}(i) \ .
\eeq
Under a gauge transformation $G \in {\cal G}$, from the transformation
law of the field $A_{\mu}$ and the definition of $U_{\mu}(i)$ given
in~(\ref{eq:link}), one immediately obtains the transformation
\beq
U_{\mu}(i) \to G(i) U_{\mu}(i) G^{\dag}(i + \hat{\mu}) \ .
\eeq
As a consequence, the parallel transport of the link variable along an
elementary square of the lattice (known as the {\em plaquette})
\beq
U_{\mu \nu}(i) = U_{\mu}(i) U_{\nu}(i + \hat{\mu}) U^{\dag}_{\mu}(i +
  \hat{\nu}) U^{\dag}_{\nu}(i) 
\eeq
transforms in the adjoint representation of ${\cal G}$:
\beq
\label{eq:gauge:plaq}
U_{\mu \nu}(i)\to G^{\dag}(i) U_{\mu \nu}(i) G(i) \ .
\eeq
On the lattice, $U_{\mu \nu}(x)$ plays the same role as the field
tensor $G_{\mu \nu}(x)$ in the continuum. 

The simplest form for the lattice action of gauge fields is given by
the Wilson action
\begin{eqnarray}
  S_W = \beta \sum_{i, \nu < \mu} \TR \left[ 1 - \RE \left(U_{\mu
        \nu}(i) \right)  \right] \ ,
\end{eqnarray}
where $\beta = 2 N/g^2$, $\TR$ indicates the trace and $\RE$
is the real part. Note that the Wilson action is written only in terms
of the trace of (the real part of) the plaquette, and, as a
consequence of Eq.~(\ref{eq:gauge:plaq}), is gauge-invariant, as it
should be. It is possible to show that in the naive continuum limit $a
\to 0$
\beq
\label{eq:guage:action}
S_W = - \frac{1}{2 g^2} \sum_{i} \left[ \TR \left( G_{\mu \nu}(ai) G^{\mu \nu}
(ai) \right) a^4 + {\cal O}(a^6) \right] \simeq S_{\mathrm{cont}} + {\cal
O}(a^2) \ ,
\eeq
i.e. the lattice action differs from the continuum action by terms
that are of order $a^2$. A weak coupling calculation shows that the
subleading terms are irrelevant in the RG sense as
the cut-off $a$ is removed. Hence, the Wilson action is in the same
universality class as the continuum action, i.e. the two actions
describe the same continuum physics. 

With the gauge part of the action formulated in terms of variables
defined on links, gauge invariance of bilinears involving fermion
fields is easy to implement. The naive discretisation of the
derivative term in the Dirac action is done in terms of finite
differences. In the lattice action, when the natural choice of
defining fermion fields on lattice sites is performed, this
determines the appearance of bilinear terms involving spinors
evaluated on nearest-neighbour lattice points, i.e. terms of the form
$\overline{\psi}(i) \psi(i + \hat{\mu})$. Here $\psi$ is the lattice
spinor, which corresponds to the continuum spinor multiplied by
$a^{3/2}$ to make it dimensionless. Moreover, for simplicity we will
start by considering fermions in the fundamental representation. In
the presence of a gauge field, the fermion bilinear terms introduced
above are modified with the insertion of the link variable living on
the lattice link joining the relevant sites:
\beq
\overline{\psi}(i) \psi(i + \hat{\mu}) \to \overline{\psi}(i)
U_{\mu}(i) \psi(i + \hat{\mu}) \ ,
\eeq
with the resulting term being gauge invariant if $\psi(i) \to G(i)
\psi(i)$ and $\overline{\psi}(i) \to  \overline{\psi}(i) G^{\dag}(i)$,
as a straightforward discretisation of gauge transformations would
suggest. 

The final building block of our lattice gauge theory is the action of
a free fermion field. On the continuum, this can be expressed as
\beq
S_f = \int \overline{\psi}(x) D(x,y) \psi(y) \mathrm{d}^4 x
\mathrm{d}^4 y \ ,
\eeq
where the Dirac operator $D$ is the (diagonal) bilinear form
\beq
D(x,y) = \delta^4(x - y) \left(i \dslash_y - m \right) \ ,
\eeq
with $m$ being the Lagrangian fermion mass. With the discretisation of
the derivative, this takes the form of a band-diagonal matrix, in which only diagonal terms and terms
connecting nearest-neighbours are different from zero. However, a
straightforward discretisation of the Dirac operator leads to the
notorious fermion doubling problem: for each lattice flavour, sixteen
degenerate flavours are generated in the continuum limit. The doubling
problem is a consequence of the Nielsen-Ninomya no-go theorem, which
states the impossibility of having a lattice action that respects
chiral symmetry, has only nearest-neighbour interactions and is free
from doublers, showing that the preservation of these three properties
is intimately related to the realisation of the Lorentz group,
explicitly broken by the lattice discretisation. Hence, in order to have a
lattice action with a continuum limit free of doublers, one needs to
give up either ultralocality (i.e. the fact that interactions have a
finite radius) or break explicitly chiral symmetry at $m = 0$. 
At this stage, it is worth stressing that regardless of the
discretisation strategy chosen, in the continuum limit one must
recover the original theory. Hence, different discretisations will be
equivalent near the continuum limit. From the computational point of
view, however, there could be advantages in choosing a formulation over
the other. Below, we focus on the Wilson formulation, which explicitly
breaks chiral symmetry at the Lagrangian level. Chiral symmetry is
then recovered by tuning the unrenormalised fermion mass to a critical
value that is itself an output of the calculation. The advantage of
this approach over others relies on smaller computational costs while
keeping the physics close to that of the continuum formulation.

In the Wilson formulation, the lattice Dirac operator in the presence
of gauge fields is given by
\beq
  D(i,j) = \delta_{i,j} 
  - \kappa \left( \left(1-\gamma_{\mu}  \right) R\left[U_{\mu} (i)
    \right]\delta_{j,i+\hat{\mu}}  + 
    \left(1+\gamma_{\mu}  \right) R\left[U^{\dagger}_{\mu} (i-\hat{\mu})
    \right] \delta_{j,i-\hat{\mu}}   \right) \ ,
\eeq
where $k = 1/(8+2 a m)$ is called the {\em hopping parameter} and we
have allowed for the fermion field to be in a generic representation
$R$ of ${\cal G}$ by indicating by $R\left[U\right]$ the element $U
\in {\cal G}$ expressed in the representation $R$. Note that when $m =
0$, $\kappa = 1/8$, which leads to explicit breaking of the chiral
symmetry. Because of this breaking, the mass gets additively
renormalised and chiral symmetry is recovered at a value $k_c$ of
the hopping parameter that needs to be determined in the
simulation.  With this definition of the Dirac operator, we can
write the lattice fermion action for $N_f$ degenerate fermion flavours
described by the spinors $\psi_1, \ \dots, \ \psi_{N_f}$ as 
\beq
S_f = \sum_{l=1}^{N_f} \sum_{i,j} \overline{\psi}_l(i) D(i,j) \psi_l(j)
\eeq
and the full action as
\beq
S = S_f + S_W \ ,
\eeq
with $S_W$ given in Eq.~(\ref{eq:guage:action}). 

After performing the Grassmann integrals over the fermion fields, the path integral of
the theory reads
\beq
Z = \int \left( {\cal D} U_{\mu}(i) \right)  \left(\rm{det} (D)
  \right)^{N_f} e^{- S_W} \ ,
\eeq
where $\rm{det}(D)$ is the determinant of the Dirac operator and
$\left( {\cal D} U_{\mu}(i) \right) $ the path integral measure of the
link variables. In the path integral formulation, the vacuum
expectation value of an operator $O \equiv O(\psi_1, \ \dots, \psi_{N_f}; \
U_{\mu}$) is given by
\beq
\label{eq:pathintegral}
O = \frac{1}{Z} \int \left({\cal D} \psi(i) \right) \left({\cal
    D}\overline{\psi}(i) \right)  \left( {\cal D} U_{\mu}(i) \right)
O(\psi_1, \ \dots, \psi_{N_f}; U_{\mu}) \ e^{ - S} \ ,
\eeq
where we have introduced the fermion path integral measure
$\left({\cal D} \psi(i) \right) \left({\cal D}\overline{\psi}(i)
\right) $. 

If the theory is formulated on a finite lattice, expressions
like~(\ref{eq:pathintegral}) give rise to well-defined integrals that
can be evaluated efficiently using Monte Carlo techniques: after
integrating out the fermion fields and mapping the gauge fields onto a
Markovian process, the latter generates configurations 
distributed according to the path integral measure weighted with the
Boltzmann term $e^{- S + N_f \log \rm{det}(D)}$, which is then interpreted as a probability
measure. Since these configurations carry already the correct
information about the probability measure, observables are determined
as simple averages over the Markovian process. With $C^{(i)} \equiv \{ U_{\mu}^{(i)}\}$ 
the realisation of the fields at step $i$ of the Markov process, we
define $O_N$ as the estimate  of $\langle O \rangle$ for a Markov
chain of length $N$:
\beq
O_N = \frac{1}{N} \sum_{i=1}^N O(C^{(i)}) \ .
\eeq
It can be proved that 
\beq
\langle O \rangle - O_N = {\cal O}(N^{-1/2}) \ ,
\eeq
i.e. $O_N$ converges to $O$ in the limit $N \to \infty$, with the
difference being of order $1/\sqrt{N}$ for finite $N$. Hence, this
approximation is controlled. Moreover, a statistical error (computed
as the standard deviation of $O$ over the probability measure) gives
the confidence interval of the measurement.

Following the ideas outlined above, in a numerical study of a lattice gauge
theory one first investigates the value of an observable $O$ at fixed
parameters $k$ and $a$ on a fixed-size grid. Then, the calculation is
repeated, extrapolating first to infinite volume, then to the chiral
limit and finally removing the cut-off $a$ (which in an asymptotically
free gauge theory amounts to sending $\beta \to \infty$). 
 
A prerequisite for the lattice formulation is the Wick rotation, to
move from Minkowki to Euclidean space. In fact, all the considerations
presented so far have been performed on a lattice with Euclidean
metric. While in principle  it is possible to reconstruct Minkowskian
correlation functions from the Euclidean ones, a wide class of
observables (e.g. spectral masses) can be accessed directly in
Euclidean space. Among the observables we are interested in, a central
role will be played by masses of bound states, which are extracted from
correlation functions of operators transforming with the quantum
numbers of interest. If $C(\tau)$ is a correlation function of
operators of quantum numbers $J^{PC}$, from general principles we know that
\beq
C(\tau) \mathop{\simeq}_{\tau \to \infty} e^{- \tau M_{J^{PC}} }  \ ,
\eeq
where $M_{J^{P C}}$ is the lowest mass in the sector with quantum numbers
$J^{PC}$. Hence, in order to extract masses, one measures expectation
values of correlators in lattice simulations and fit the expected
asymptotic behaviour. 

Since in the Wilson formulation of lattice fermions the Lagrangian mass is
additively renormalised, it is useful to define a mass that is only
multiplicatively renormalised. This can be done through the axial Ward
identity. The corresponding mass, $\mpcac$, is zero in the chiral
limit, and can hence be used to understand how the massive case
approaches the massless limit. For this reason, we shall use $\mpcac$
instead of the Lagrangian fermion mass in our analysis in the
following section.  As a function of $\mpcac$, we can
determine the anomalous dimension from the spectrum of the theory
using the relation provided in Eq.~(\ref{eq:scaling}). When dealing
with a lattice of finite size, it is important to keep into account
the finite extension $L$ in the calculation of scaling behaviours. As a
result~\cite{Lucini:2009an}, Eq.~(\ref{eq:scaling}) is replaced by
\beq
L a m_X = F_X\left( (a L \mpcac)^{\frac{1}{1+\gamma_{\ast}}}\right) \ ,
\eeq
where $F_X$ is a universal function of the scaling variable $x = L (a 
\mpcac)^{\frac{1}{1+\gamma_{\ast}}}$. The practical meaning of this
expression is that values of $m_X$ at
different lattice sizes $L$ are described by a universal curve of
$x$. Due to striking similarities with Monte Carlo investigations of
second order phase transitions (which IR conformality is closely
related to), this method is called Finite Size Scaling (FSS). 

Having introduced the generalities of Lattice Gauge Theories, Monte
Carlo simulations and the analysis techniques we shall be using to
understand the IR conformal regime, below we examine some
non-perturbative results for gauge theories that are relevant for our
understanding of strongly interacting dynamics beyond the SM,
using as a guide the discussion of Sect.~\ref{sect:2}.

\section{SU(2) Gauge Theory with two adjoint Dirac flavours}
\label{sect:4}
One of the first theories potentially relevant for dynamical
electroweak symmetry breaking~\cite{Dietrich:2006cm} that was studied numerically is the
SU(2) gauge theory with two adjoint fermions. After the first
explorations~\cite{Catterall:2007yx,DelDebbio:2008zf,Catterall:2008qk,Hietanen:2008mr},
systematic investigations of the spectrum were
performed~\cite{DelDebbio:2009fd,Catterall:2009sb,DelDebbio:2010hx,DelDebbio:2010hu},
which pointed out how the general behaviour of bound state masses as a
function of the fermion mass is qualitatively different from the one
observed in QCD. In particular, it was noticed
in~\cite{DelDebbio:2009fd} and confirmed with a more detailed analysis
in~\cite{DelDebbio:2010hx,DelDebbio:2010hu}, that the behaviour of the
spectrum in the chiral limit is compatible with the scenario presented
in~\cite{Miransky:1998dh}, where in the vicinity of a weakly coupled
IR fixed point a dynamical Yang-Mills scale is generated that is
exponentially suppressed with respect to the renormalised fermion mass.  

\begin{figure}
\begin{center}
\includegraphics[width=0.5\textwidth]{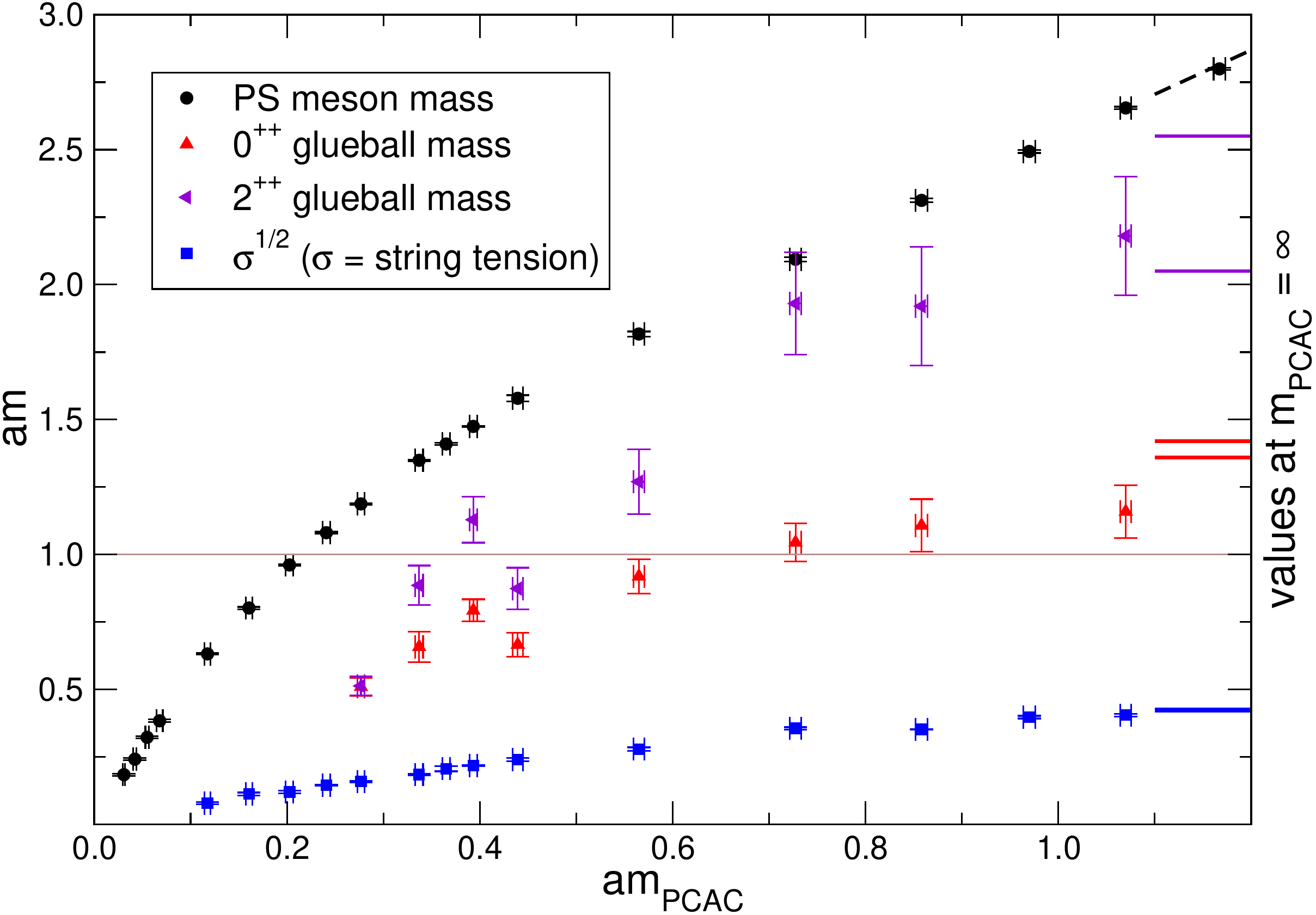}
\end{center}
\caption{The spectrum of SU(2) gauge theory with two Dirac adjoint
  flavours as a function of the PCAC mass at $\beta = 2.2$
  (from~\cite{DelDebbio:2009fd}).}
\label{fig:all}
\end{figure}
The results of an investigation of the spectrum of the theory are
plotted in Fig.~\ref{fig:all}. This is reminiscent of the scenario of
locking when the locking scale is much larger than the  Yang-Mills
scale (see~Fig.~\ref{fig:lock}b). Once again, this is compatible
with the IR fixed point being of the Caswell-Banks-Zaks type, which
determines the emergence of a spectrum that shares the qualitative
features of that of a QCD-like theory with heavy quarks. Note that this
spectrum arises in the IR independently of the Lagrangian
fermion mass, which, although in a non-trivial way, in this case plays
the sole role of setting the scale of the effective IR Yang-Mills
theory. Another indication pointing in this direction is the
degeneracy of the pseudoscalar meson and the vector
meson~\cite{DelDebbio:2009fd,DelDebbio:2010hu}. 

\begin{table}
\begin{center}
\begin{tabular}{|c|c|} \hline \hline
Method & $\gamma_{\ast}$ \\ \hline \hline
FSS~\cite{Lucini:2009an} & $0.05 < \gamma_{\ast} < 0.25$ \\ \hline 
SF~\cite{Bursa:2009we} & $0.05 < \gamma_{\ast} < 0.56$ \\ \hline 
FSS~\cite{DelDebbio:2010hu} & $0.05 < \gamma_{\ast} < 0.20$ \\ \hline
FSS~\cite{DelDebbio:2010hx} & $0.22 \pm 0.06$ \\ \hline
MCRG~\cite{Catterall:2011zf} & $-0.6 < \gamma_{\ast} < 0.6$ \\ \hline
SF~\cite{DeGrand:2011qd} & $0.31 \pm 0.06$ \\ \hline 
FSS~\cite{Giedt:2012rj} & $0.51 \pm 0.16$ \\ \hline 
MNS~\cite{Patella:2012da} & $0.37 \pm 0.02$ \\ \hline 
MNS~\cite{Debbio:2014wpa}& $0.38 \pm 0.02$ \\ \hline \hline
Perturbative 4-loop~\cite{Pica:2010xq} & $0.500$ \\ \hline
All-orders hypothesis~\cite{Pica:2010mt} & $0.46$ \\ \hline \hline
\end{tabular}
\end{center}
\caption{Summary of numerical results for the anomalous dimension
  $\gamma_{\ast}$ in SU(2) gauge theory with two adjoint Dirac
  fermions. For comparison, some analytical estimates are reported in
  the last two lines. For details about methods and analyses, we
  refer to the quoted literature.}
\label{tab:1}
\end{table} 

From a more quantitative perspective, precise numerical data allow us to
extract the value of the anomalous dimension of the chiral
condensate. The first measurement for this theory was provided
in~\cite{Lucini:2009an} using a FSS technique. Subsequently, other
analyses using this and alternative methods have been performed. A
summary of results obtained using FSS, the Schr\"odinger Functional
(SF) technique, Monte Carlo Renormalisation Group (MCRG) methods and
the Dirac Mode Number Scaling (MNS) is provided in Tab.~\ref{tab:1},
together with reference to the original works that describe in details
those methods and the corresponding results.
Within the variability of the results (for which the most recent ones
have to be considered as more reliable, since they better reflect the
evolution of our understanding of the underlying tools and
techniques), the clear feature that emerges is that the anomalous
dimension appears to be well below one, in stark contrast with the
phenomenological requirements. Hence, the conclusion that can be drown
is that, while displaying an interesting IR behaviour that is
clearly distinguished from that of QCD-like theories and in principle
promising as a mechanism of dynamical electroweak symmetry breaking,
from a quantitative point of view the model is not phenomenologically
viable. The small anomalous dimension is compatible with the
qualitative observation that the emerging Yang-Mills spectrum has a
constituent quark mass well above the Yang-Mills scale, a feature that
characterises IR conformal theories with a fixed point of the
Caswell-Banks-Zaks type~\cite{Miransky:1998dh}. Despite the small anomalous dimension, the model has
proven to be very interesting from the conceptual point of view, as it
shows at work some of the key ideas of technicolor. 

To our knowledge, SU(2) with two adjoint Dirac fermions is the first gauge theory that has been shown to
display an IR conformal or near-conformal behaviour, the two
scenarios being very hard to disentangle in a lattice
calculation\footnote{For this reason, exceeding on the side of caution, we
prefer to talk about (near-)conformal behaviour, indicating with this
expression that both conformality and near-conformality are in fact
possible.}. Over the years, the model has been object of wide
attention and more refined
studies~\cite{Karavirta:2011mv,Bursa:2011ru,Karavirta:2012qd,Bennett:2012ch,Rantaharju:2013gz,Rantaharju:2013bva}. 
Currently, there seems to be no doubt in the field that the IR
behaviour of the theory is 
definitely different from that of QCD. Despite the little relevance for
phenomenology, the theory is still the subject of intensive numerical
study, since it can serve as a solid toy model to understand
(near-)conformal gauge theories. Ongoing
investigations~\cite{Debbio:2014wpa} are focussing on a better understanding
of finite size effects, which play a key role in numerical studies of
IR conformal systems. The first indications seem to confirm the
broad picture that emerged on smaller lattices. The next step will be
to perform a systematic exploration towards the continuum limit. This
could help to understand whether the differences in the extracted values of the
anomalous dimension of the condensate could be due to scaling
violations related to the fact that the gauge coupling is marginal in
the RG sense~\cite{Cheng:2013xha}. 

\section{SU(2) Gauge Theory with one adjoint Dirac flavour}
\label{sect:5}
The model discussed in the previous section shows at work some of
the ideas that motivate us to study strongly interacting dynamics beyond
the SM as a mechanism of electroweak symmetry
breaking. However, it fails the crucial test of phenomenological
viability, due to the small anomalous dimension of the
condensate. From a theoretical point of view, we can ask ourselves
whether models with an anomalous dimension that is compatible with
phenomenological requests do exist. The general agreement is that anomalous
dimensions of order one can arise near the onset of the conformal
window.  In the case of SU(2) with adjoint Dirac fermions, given that the
case $N_f = 2$ is (near-)conformal, we can ask what happens when we
take one single flavour. A possibility is that the theory be
confining in a QCD-like fashion. However, it is also possible that it
be walking or stay conformal. If the IR dynamics is different
from that of QCD, we expect to observe an anomalous dimension that is
larger than that of the two flavour case and, if the theory is near the
onset of conformality, the anomalous dimension should be of order
one. 

\begin{figure}
  \begin{tabular}{cc}
    \includegraphics[width=0.45\textwidth]{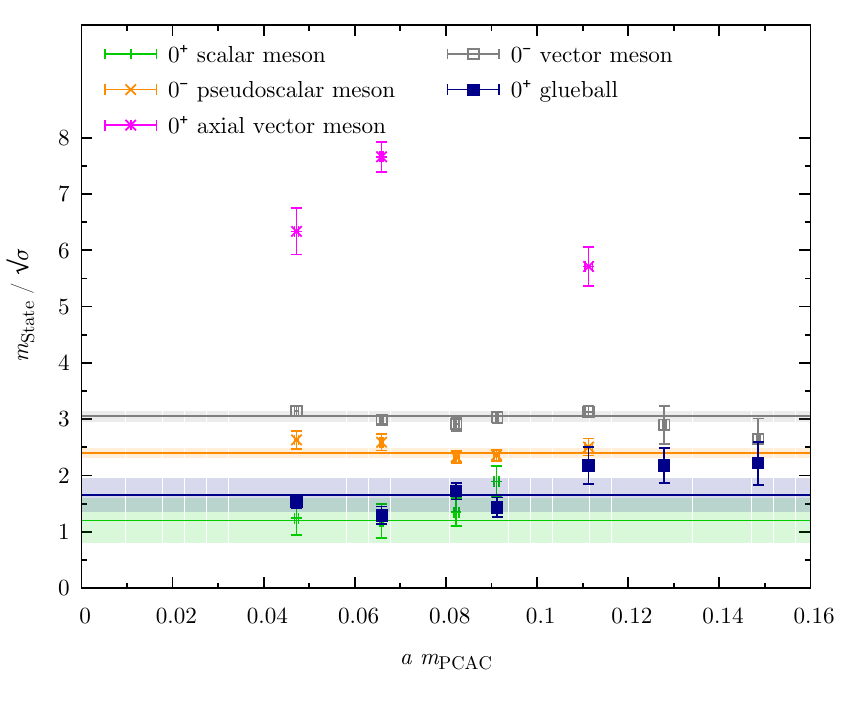} &
    \includegraphics[width=0.45\textwidth]{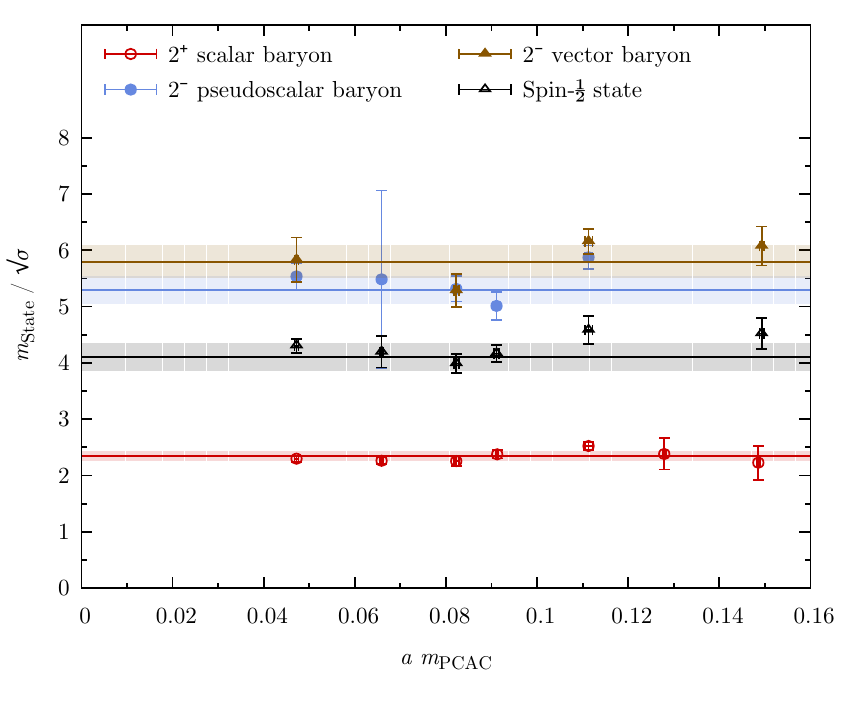}    \\
    (a) & (b)
  \end{tabular}
  \caption{Spectrum of the theory, showing (a) meson and
    glueball, and (b) baryons and \spinhalf states, all
    normalized by $\sigma^{1/2}$. From~\cite{Athenodorou:2014eua}. \label{fig:spectrum}} 
\end{figure}

We can test these expectations by performing numerical simulations of
the one-flavour model~\cite{Athenodorou:2013eaa,Athenodorou:2014eua}.  We focus
our attention on the mass spectrum of the theory at finite fermion
mass, aiming at extrapolating to the chiral limit to probe the
IR behaviour of the theory for massless constituent
fermions. We first note that the theory has a non-trivial chiral
symmetry breaking pattern that reduces the global symmetry from SU(2)
to a residual SO(2). This chiral symmetry breaking pattern can be
exposed by rewriting the Dirac flavour in terms of two Majorana or two
Weyl components. We refer
to~\cite{Athenodorou:2013eaa,Athenodorou:2014eua} for details. The
classification of fermion bilinears in terms of their transformation
law under the Lorentz group, parity and the residual SO(2) group
(which leads to the conservation of a quantum number that can be taken
as the baryonic charge) shows that the spectrum contains mesons
(i.e. bound states of two fermions  with zero baryon number), baryons
(or diquarks), which are bound states of two fermions with baryon
number $B = \pm2$ and \spinhalf states (single fermions dressed with
gluons) in addition to the usual Yang-Mills glueball spectrum. The
Goldstone bosons emerging from the spontaneous breaking of the chiral
symmetry are scalar baryons. For a full classification of the
spectrum, we refer to~\cite{Athenodorou:2014eua}. 

The spectrum of the theory is reported in
Figs.~\ref{fig:spectrum}a~and~\ref{fig:spectrum}b in units of the
square root of the string tension $\sigma$. These figures show that mass ratios
are constant towards the massless limit. As previously seen, this is a
typical signature of IR (near-)conformality. We also note that
the scalar baryon (shown in Fig.~\ref{fig:spectrum}b) is heavier than
the scalar meson, which in turn has the same mass as the scalar
glueball (Fig.~\ref{fig:spectrum}a). The would-be Goldstone boson being
not the lightest state of the theory when the mass goes to zero is
incompatible with the scenario of chiral symmetry breaking, while it is
possible in an IR (near-)conformal gauge theory. As already
noticed, the scalar glueball and the scalar meson are degenerate. This
indicates that those states (which are notionally distinct in
QCD) are in fact the same state, which emerges as the lightest scalar
state either by computing scalar correlators in the mesonic sector or
by computing correlators of gluonic operators that are invariant under
spatial rotations and parity. 

\begin{figure}
  \begin{tabular}{cc}
    \includegraphics[width=0.45\textwidth]{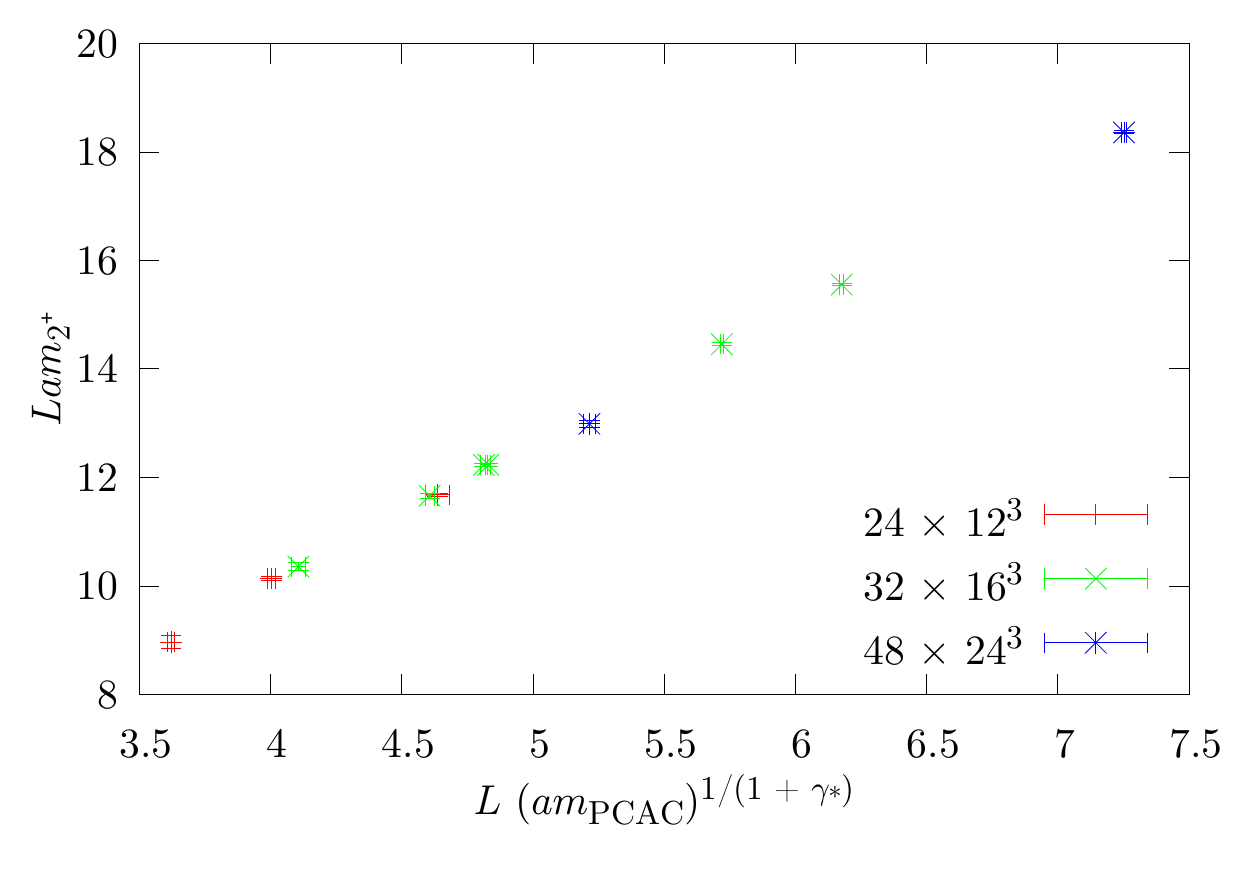} &
    \includegraphics[width=0.52\textwidth]{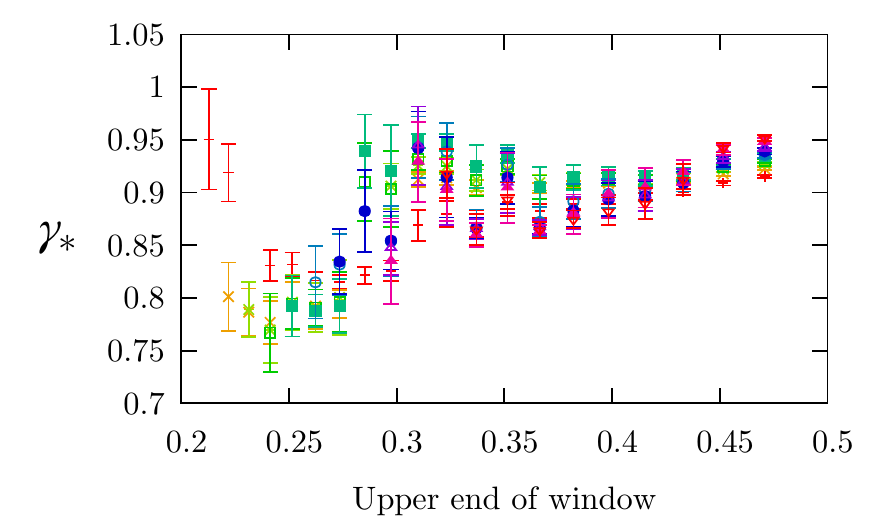} \\
    (a) & (b)
  \end{tabular}
  \caption{(a) The mass of the $B = 2$ baryon after rescaling of the
    data assuming $\gamma_{\ast} = 1$; (b) anomalous dimension
    $\gamma_{\ast}$ determined by studying the Dirac operator
    mode number scaling in a calculation performed on a $48 \times
    24^3$ lattice at $a m = 1.523$ and $\beta = 2.05$.
    From~\cite{Athenodorou:2014eua}. \label{fig:gammastar}}  
\end{figure}

A FSS analysis of the spectrum can be performed to get a handle on the
anomalous dimension of the condensate. In Fig.~\ref{fig:gammastar}a we
plot the combination $a L m_{2^+}$ (with $m_{2^+}$ the mass of the
$B = 2$ baryon measured on a lattice of extension $2L \times L^3$,
where $2 L$ is the temporal size and $L$ the spatial
size) as a function of $L a \mpcac^{1/(1  + \gamma_{\ast})}$ assuming
  $\gamma_{\ast} = 1$. The collapse of the data onto a single 
curve shows that the anomalous dimension takes values near
one. Inspection of similar plots obtained by varying
$\gamma_{\ast}$ suggests that $0.9 \le \gamma_\ast \le 1$. The anomalous
dimension can be studied also using the MNS
technique described in~\cite{Patella:2012da}. Fig.~\ref{fig:gammastar}b shows 
a typical results of fits of the anomalous dimension when varying the
fit interval (whose upper value is the abscissa of the plot and
various lower values are represented by the different points at fixed
abscissa). For reasonable values of the upper value (between 0.35 and
0.45) the fit gives stable results that can be summarised as
$\gamma_\ast = 0.925(25)$, a value that is compatible with that
extracted using FSS techniques. 

The lattice results reported in this section (which should still be
regarded as exploratory, as simulations have been performed for a
single $\beta$ on a limited set of volumes and fermion masses)
suggest that (a) SU(2) gauge theory with one adjoint Dirac flavour is
(near-)conformal; (b) the theory has an anomalous dimension of the
order of that required by phenomenology; (c) the lightest particle is
the scalar particle. From the conceptual point of view, these three
properties are very important, because they show from first principles
that near the conformal window anomalous dimensions of order one do
arise and the scalar (i.e. the would be Higgs particle of the SM)
is naturally light. Although this theory is still of no
phenomenological interest, as the symmetry breaking pattern can
account for the mass of only two SM gauge bosons, our results provide a
numerical proof that the key ideas of strongly interacting dynamics
beyond the SM are realised in concrete systems.

\section{Conclusions}
\label{sect:6}
In this work, we have presented lattice results aimed at answering
some of the most crucial conceptual questions still open in the
framework of strongly interacting dynamics beyond the SM. We have
provided numerical evidence of the existence of 
IR (near-)conformal gauge theories by anticipating their
spectral signature and finding confirmation for it in candidate
theories using first-principle Monte Carlo calculations. The
investigation of SU(2) gauge theory with one adjoint Dirac fermion
suggests that there exist  strongly interacting theories that are
(near-)conformal and have an anomalous dimension of order one and a
light scalar particle. The large anomalous dimension is needed to fulfil
the phenomenological constraints provided by electroweak precision
measurements, while the light scalar is identified with
the Higgs boson. The fact that LHC experiments have not seen other
new particles in addition to the Higgs is compatible with the scalar being the lightest particle
in the spectrum of the novel strong dynamics. With the LHC restarting
its operations, the quest for a dynamical explanation of
electroweak symmetry breaking might soon find an answer. While none of the models
studied so far can provide a realistic phenomenological description
of electroweak symmetry breaking due to a new strong force, lattice
investigations (which include studies of other theories that we have
been unable to cover here, see e.g.~\cite{Appelquist:2007hu,Deuzeman:2008sc,Shamir:2008pb} and more
recently ~\cite{Lin:2012iw,Aoki:2013zsa,Fodor:2014cpa,Weinberg:2014ega,Lombardo:2014fea} for
other approaches and results in different models) have greatly
contributed to a deeper understanding of 
crucial non-perturbative features of the framework. Together with other
conceptual advances, they have confirmed strongly interacting dynamics
beyond the SM as one of the most robust sets of ideas that
could explain electroweak symmetry breaking~\cite{Arbey:2015exa}.

\section*{Acknowledgments}
It is a pleasure to thank A. Athenodorou, E. Bennett, G. Bergner,
F. Bursa, L. Del Debbio, D. Henty, E. Kerrane, A. Patella, T. Pickup,
C. Pica, A. Rago, E. Rinaldi, R. Sabin for their valuable
contributions to the collaborations on which this work is
based. Numerical computations were executed in part on the Blue Gene/P
machine in Swansea University and the ULGQCD cluster in the
University of Liverpool (part of the DiRAC facility supported by
STFC), on the HPC Wales cluster in Cardiff, supported by the ERDF
through the WEFO (part of the Welsh Government), on the BlueGene/Q
system at the Hartree Centre (supported by STFC) and on the BlueGene/Q
system at the University of Edinburgh (part of the DiRAC2 facility
supported by STFC). This work has been supported by STFC (grant
ST/G000506/1). 

\section*{References}
\bibliographystyle{iopart-num}
\bibliography{references}

\end{document}